\documentclass[9pt,twocolumn,twoside]{osajnl}

\journal{ol} 

\setboolean{shortarticle}{true}

\usepackage{lineno}

\title{Passive repetition-rate stabilization for a mode-locked fiber laser by electro-optic modulation}

\author[1]{Tingting Yu}
\author[1]{Shuhong Jiang}
\author[1]{Jianan Fang}
\author[1]{Tingting Liu}
\author[1]{Xiuqi Wu}
\author[1,2]{Ming Yan}
\author[1,2,3,*]{Kun Huang}
\author[1,2,4]{Heping Zeng}

\affil[1]{State Key Laboratory of Precision Spectroscopy, East China Normal University, Shanghai 200062, China}
\affil[2]{Chongqing Key Laboratory of Precision Optics, Chongqing Institute of East China Normal University, Chongqing, China}
\affil[3]{Collaborative Innovation Center of Extreme Optics, Shanxi University, Taiyuan, Shanxi 030006, China}
\affil[4]{Jinan Institute of Quantum Technology, Jinan, Shandong 250101, China}
\affil[*]{khuang@lps.ecnu.edu.cn}

\begin{abstract}
We report a passive stabilization of the repetition rate for a mode-locked fiber laser by using an electro-optic modulator in a phase-biased nonlinear amplifying loop mirror. The underlying mechanism, in contrast to active feedback operations, lies in the cross-phase modulation between electrical and optical pulses within an electro-optic crystal. The resulting spectral shift can automatically compensate the cavity-length drift via the group velocity dispersion. Consequently, the artificial actuator enables to obtain a capture range up to 2.3 mm, much longer than that achieved by index changes of the modulator.  A robust and tight locking for the repetition rate is then realized with a standard deviation as low as 9 $\mu$Hz with a 1-s sample time over 11 hours, corresponding to a fractional instability of 4.3$\times$10$^{-13}$. Furthermore, a dynamic optical sampling by repetition-rate tuning has been manifested with a fast refresh rate at 100 kHz and a broad scanning range over 305 ps. The demonstrated passive servo action may provide a simple yet effective way to stabilize the repetition rate with high precision, large bandwidth and wide tunability.
\end{abstract}

\setboolean{displaycopyright}{true}

\begin{document}

\maketitle

Mode-locked fiber lasers under spectro-temporal control have become prevalent tools in a variety of applications \cite{Kim2016AOP}, such as  optical communication, frequency metrology and precision ranging. In particular, the stabilization of the repetition rate is essential to implement optical frequency combs, thus favoring precise timing of pulse trains and accurate positioning of spectral modes \cite{Diddams2020Science}. Up to date, there have been various techniques proposed to stabilize the round-trip time of optical pulses within the laser cavity \cite{Yang2018LP}. The most common method is to precisely control the geometric cavity length by using a servo-controlled piezoelectric transducer (PZT). However, the feedback bandwidth of the PZT actuator is typically limited to the few kHz-level due to strong mechanical resonances \cite{Zhang2009OE}. Alternatively, the round-trip optical path could be stabilized by modulating the pump power on a gain medium \cite{Rieger2013OE, Hao2016JLT} or controlling the polarized state of propagation pulses \cite{Shen2015APL}. Similarly, the servo bandwidth is also restricted to kHz due to the long upper-state lifetime or limited polarization switching speed. 

In this context, an intracavity electro-optic modulator (EOM) has been used to provide a broad-bandwidth feedback loop to stabilize the repetition rate \cite{Hudson2005OL}. Consequently, the effective suppression of high-frequency noises allows to obtain a precise timing synchronization \cite{Chen2015SR} or a narrow comb linewidth \cite{Iwakuni2012OE}. One remaining inconvenience lies in the small dynamic range of the non-mechanical actuator. For instance, Ref. \cite{Hudson2005OL} reports a maximum length variation about 40 nm for a 2-cm long electro-optic crystal. As a result, a long-term operation usually requires additional actuated elements to enlarge the capture range of the locking system \cite{Kuse2016OE,Kwon2017OL}. Notably, the frequency response of the PZT-based actuator has recently been extended up to 500 kHz \cite{Nakamura2020OE} by optimizing the mechanical design and electronic circuits, yet the compensated range for the length fluctuation is still at the $\mu$m level. It is thus desirable to obtain a servo functionality with a high bandwidth and a large dynamic range, especially pertinent to field applications with alleviated requirements for the temperature stabilization and vibration isolation \cite{Yang2016AO}.

\begin{figure*}[t!]
\centering
\includegraphics[width=0.7\textwidth]{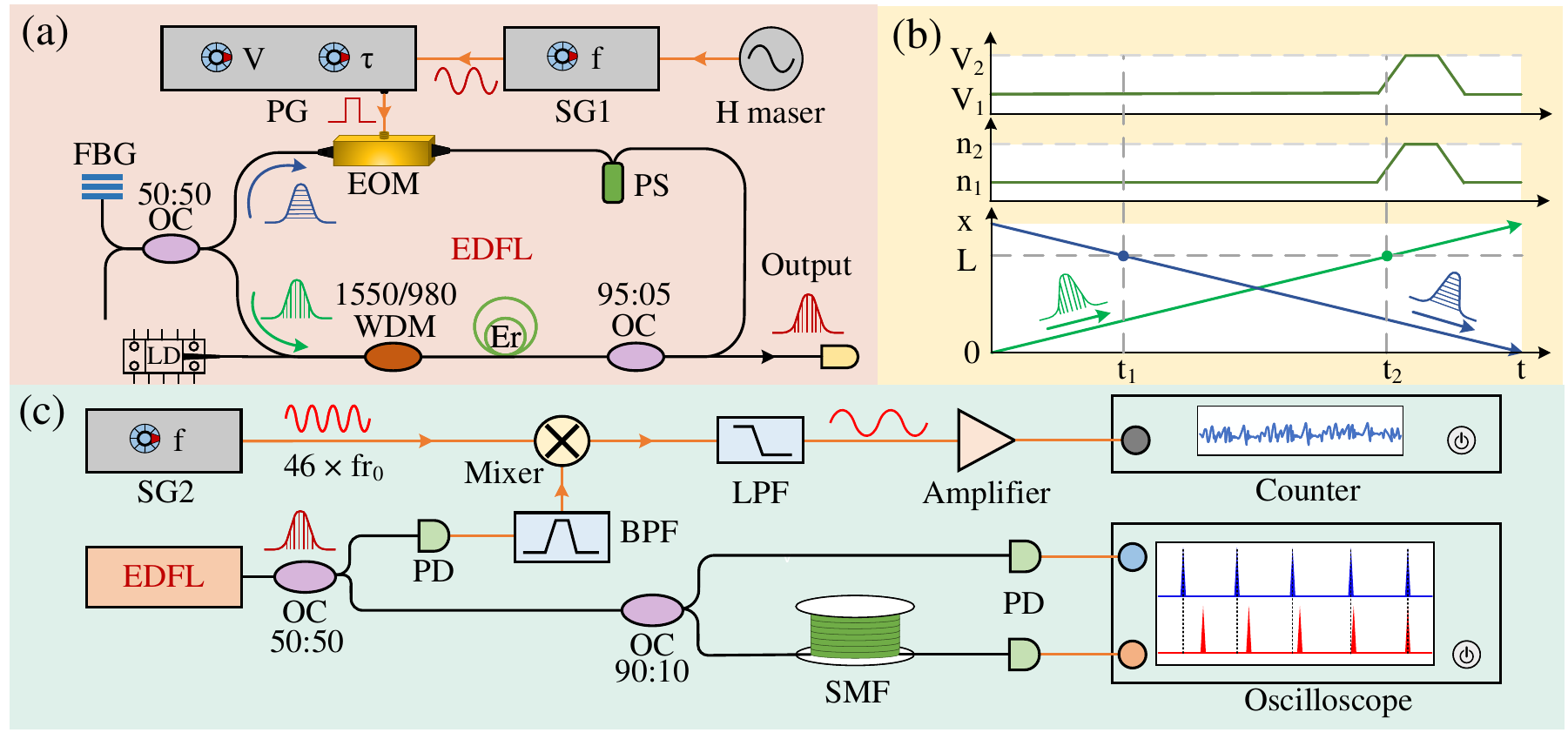}
\caption{(a) Schematic of the mode-locked fiber laser at a passively stabilized repetition rate. The erbium-doped fiber laser (EDFL) contains a waveguide electro-optic modulator (EOM), which is placed asymmetrically within the nonlinear amplifying loop mirror. The intracavity modulator is driven by a train of stable electrical pulses referenced to a stable hydrogen frequency standard. The involved interaction between the electrical and optical pulses in the electro-optic crystal would result in a repetition-rate-stabilized mode-locking state. (b) Space-time diagram for two counter-propagating pulses within the Sagnac loop. The origin is set at the OC port, while the EOM is placed at the position L. The two counter-propagating optical pulses within the Sagnac loop will pass through the EOM at $t_1$ and $t_2$. The voltage variation induces the index change of the EOM, thus imposing a non-reciprocal phase shift on the circulating optical pulses. (c) Setup for characterizing the frequency-locked laser source. One part of the EDFL output is mixed with the signal at the 46th harmonic frequency. The resulting beat signal is then sent to a high-precision counter to measure the repetition-rate fluctuation. The other part of the laser output is steered into two arms with a large delay about 780 m. The two self-delayed pulse trains are then recorded by a high-speed digital oscilloscope to demonstrate the dynamical optical sampling with a high speed and a broad range. PG: electrical pulse generator; SG: signal generator; LD: Laser diode; WDM: wavelength division multiplexer; Er: erbium-doped gain fiber; OC: output coupler; FBG: fiber Bragg grating; PS: phase shifter; SMF: single-mode fiber; PD: photodiode; BPF and LPF: band- \& low-pass filter.}
\label{fig1}
\end{figure*}

In parallel, the all-optical passive approach recently attracts increasing attention to control the intracavity pulse dynamics of ultrafast lasers, which has been demonstrated in various mode-locking operations based on nonlinear polarization rotation \cite{Yoshitomi2006, Tsai2013}, saturable absorbers \cite{Zhang2011, Li2020OL}, and nonlinear amplifying loop mirrors (NALM) \cite{Zeng2019OL, Yu2021OE}. Specifically, the optical injection into the laser cavity induces a pulling effect between master and slave pulses through the cross-phase modulation (XPM) \cite{Wei2002APB, Kim2021SA}. The resulting spectral shift is self-adapted to compensate cavity-length fluctuations due to the group-velocity dispersion, which enables to reach a large capture range up to centimeter level \cite{Zeng2019OL}. In this scenario, the high-speed response is permitted by the nearly instantaneous optical nonlinearity, thus avoiding the stringent requirement for high-bandwidth actuators and electronics in the active configuration. To lock the slave optical cavity, a master laser source is typically needed at a well-matched and stabilized repetition rate \cite{Yu2021OE, Kuse2012OE}, which inevitably adds additional complexity and cost of the system.

In this letter, we devise and implement a novel scheme to realize a passive repetition-rate stabilization for a polarization-maintaining mode-locked fiber laser. In contrast to the reported optical-injection technique, the frequency-controlled laser here is directly obtained by an electrically driven EOM within a NALM of the laser cavity. Thanks to the nonlinear XPM effect in the electro-optic crystal, the electrical signal referenced to an external clock allows to initiate and retain a synchronous mode-locking. Consequently, the capture range unprecedentedly reaches to 2.3 mm, which enables a long-term operation over 11 hours without thermal and vibration isolation. Meanwhile, the achieved standard deviation is as low as 9 $\mu$Hz with a 1-s sample time, corresponding to a fluctuation instability of 4.3$\times$10$^{-13}$. A large feedback bandwidth is further manifested with a fast tuning speed at 100 kHz over a repetition-rate change of 2 kHz, which facilitates an optical sampling range up to 305 ps between two self-delayed pulse trains. The presented passive servo with simple, compact and robust features would promote subsequent applications beyond laboratory operations.

Figure \ref{fig1}(a) presents the experimental setup for implementing the repetition-rate-stabilized fiber laser. The Er-doped fiber laser (EDFL) is mode locked based on a NALM as the artificial absorber, which outputs a pulse train at a fundamental frequency $fr_{0}$ of 21.8 MHz. In the Sagnac loop, the gain fiber is placed asymmetrically relative to the 2$\times$2 optical coupler, which favors to accumulate sufficient phase shift difference between two counter-circulating fields. Additionally, an integrated $\pi$/2 phase shifter is used to reduce the mode-locking threshold, which enables to achieve a self-starting mode-locked operation \cite{Kuse2016OE}. The core component in our laser resonator is a waveguide EOM (Keyang Photoncis, KY-PM-15-10G) with a bandwidth of 10 GHz, which is electrically driven by a pulse generator (Agilent, 81130A) to implement the phase modulation. The electrical pulses are externally triggered by a signal generator (Rohde \& Schwarz, SMC100A). The involved clock for these devices can be traced to an active hydrogen maser (Microsemi, MHM2010) with a specified instability of 1.5$\times$10$^{-13}$ in 1-s sample time. As shown in Fig. \ref{fig1}(b), the two counter-propagating pulses within the NALM pass through the EOM at different time, which results in a periodic introduction of nonreciprocal phase difference. Consequently, the electrical pulse serves as a stable and precise timing trigger for an effective intensity modulator \cite{Yu2021OE}. The repetition rate of the fiber laser will thus be identical to the driving frequency to obtain a stable mode-locking operation. Moreover, a slight spectral shift of the laser pulse can be self-adapted to stabilize the cavity-length variations, as verified in subsequent experimental results. The underlying mechanism is fundamentally different from the method relying on the refractive index change of the EOM \cite{Hudson2005OL}. Figure \ref{fig1}(c) illustrates the schematic for characterizing the interval-stabilized pulses. The frequency instability is measured by a high-resolution frequency counter (Tektronix, FCA3100). To improve the measurement precision, the beat signal around the 46th harmonic of the fundamental repetition rate is recorded. Additionally, the repetition-rate is rapidly scanned to implement an optical sampling between two delayed pulses from a single laser \cite{Hochrein2010OE}, validating a large dynamic range and a fast tuning speed for the locking system.

\begin{figure}[b!]
\centering
\includegraphics[width=0.92\columnwidth]{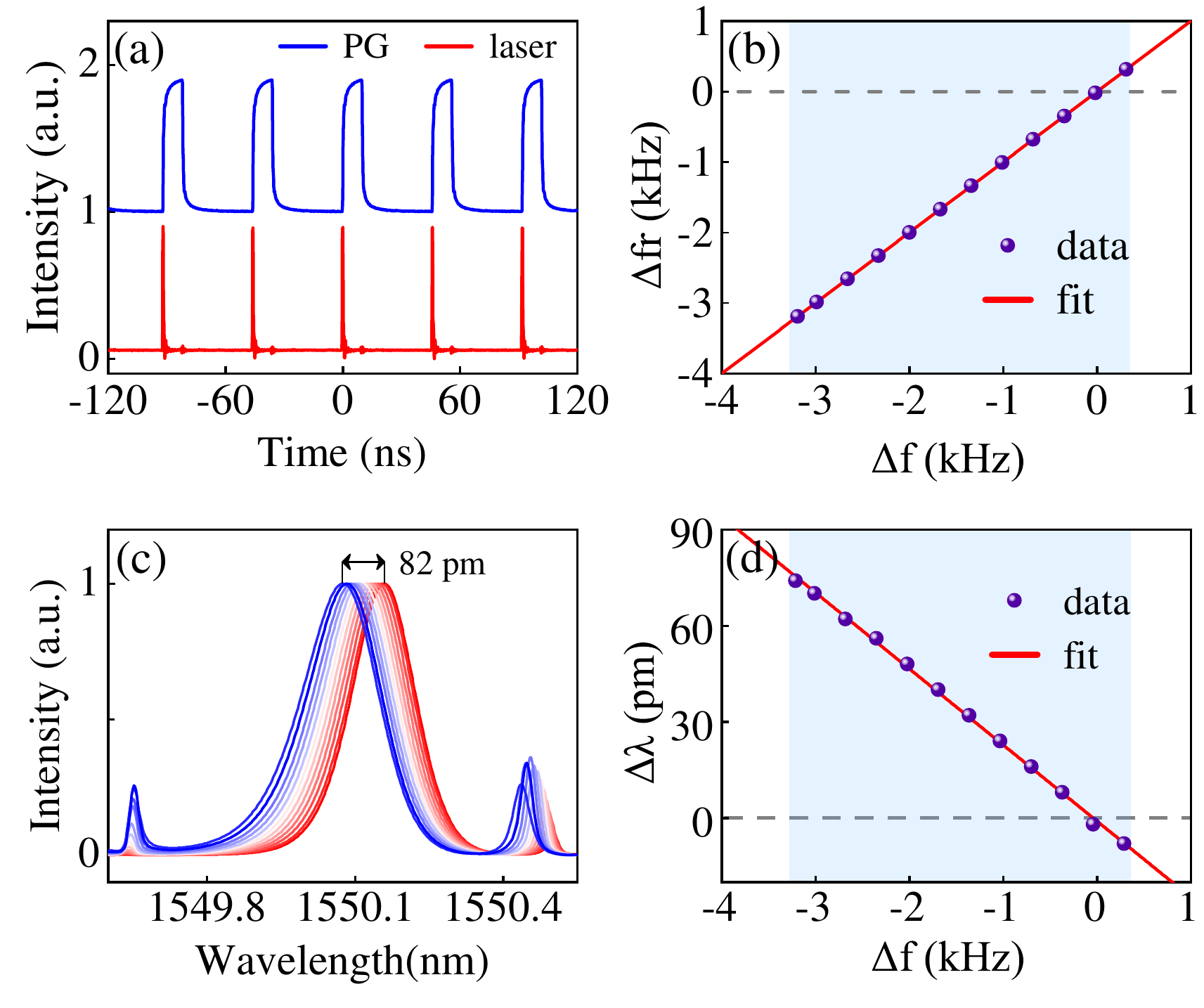}
\caption{(a) Synchronous pulse trains from the pulse generator (PG) and fiber laser. (b) Repetition rate of the fiber laser increases with the driving frequency of the electric signal. Note that the repetition rates are offset by the free-running value at 21.8 MHz. The shaded area indicates the capture range of the locking system. (c) Optical spectra at various operation frequencies for the stabilized mode-locked fiber laser. (d) Wavelength shift as detuning the repetition rate of the laser.}
\label{fig2}
\end{figure}

We first examine the passive synchronization performance between the electrical and optical pulses. To this end, the signal frequency $f$ of the SG is tuned to be near the repetition rate $fr_0$ of the fiber laser. As shown in Fig. \ref{fig2}(a), a stable trace for the optical pulse can be observed in the oscilloscope triggered by the PG pulse. The optical pulse duration is about 17 ps, closed to the Fourier limit of the 0.3-nm spectrum determined by the intracavity fiber Bragg grating (FBG). Note that the EDFL itself can be passively mode-locked in a self-starting fashion without the presence of the electrical modulation. This behavior is fundamentally different from that in the typical active mode-locking regime. Here, the relative timing jitter between the electrical and optical pulses is minimized due to the XPM within the nonlinear electro-optic crystal, which results in a passive stabilization of the laser repetition rate. The amplitude and width of the electrical pulse are set to be 1 V and 10 ns, respectively. The synchronous behavior can be seen from the one-to-one correspondence between $\Delta fr$ and $\Delta f$ in Fig. \ref{fig2}(b). The capture range is found to be about 3.7 kHz, corresponding to a free-space cavity variation of 2.3 mm. As a result, the relative change $\Delta fr / fr_0$ reaches to 1.7$\times$10$^{-4}$, which is about three orders of magnitude larger than the reported value of 1.3$\times$10$^{-7}$ \cite{Hudson2005OL}. Figure \ref{fig2}(c) presents the output optical spectrum as varying the driving frequency. A central wavelength shift of 82 pm is indicated in Fig. \ref{fig2}(d). It is the induced spectral adaption that compensates the cavity-length variation according to the group velocity dispersion \cite{Yoshitomi2006, Yu2021OE}. The observed large group delay might be ascribed to the pronounced dispersion for the narrow-band FBG \cite{Qian2009OE}. Similar to the XPM between optical pulses, the involved interaction in an electro-optic medium would induce a pulling effect between the electrical and optical pulses, which leads to a round-trip self-synchronization process \cite{Wei2002APB}. The use of a larger-bandwidth FBG can support a wider wavelength shift, thus improving the capture range \cite{Zeng2019OL}.

\begin{figure}[t!]
\centering
\includegraphics[width=0.85\columnwidth]{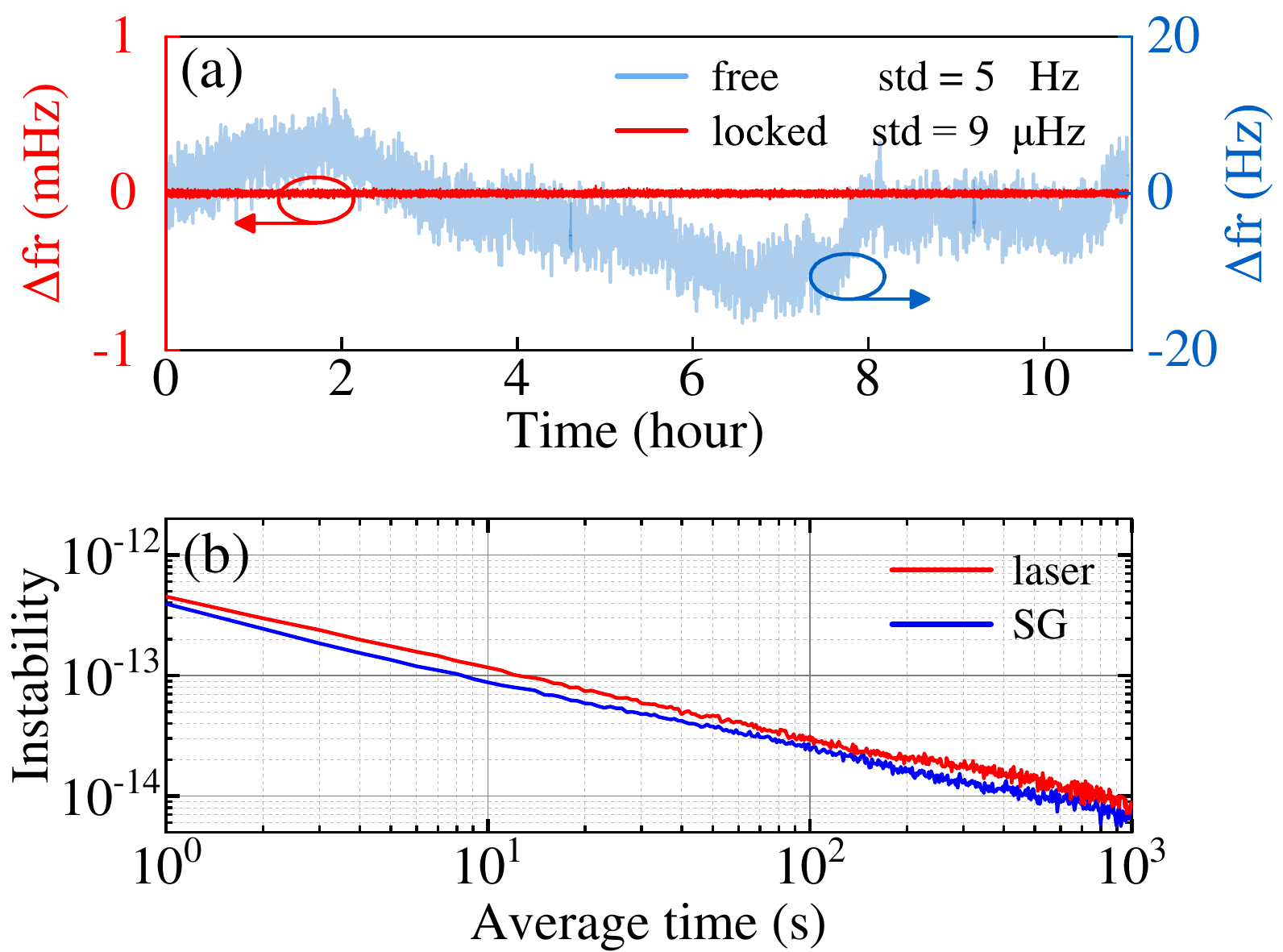}
\caption{(a) Evolution of the repetition-rate variation over 11 hours for the fiber laser at the locked and free-running operation, which indicates a standard deviation (std) of 9 $\mu$Hz and 5 Hz, respectively. The sample time is set to be 1 s. (b) Fractional instability of the repetition rate for the stabilized laser and signal generator at different averaging time.}
\label{fig3}
\end{figure}

Figure \ref{fig3} shows the long-term performance of the passively stabilized fiber laser over 11 hours, which exhibits a standard deviation of 9 $\mu$Hz and 5 Hz at the locked and free-running operations, respectively. The Allan deviation is further calculated as a function of the average time to investigate the repetition-rate instability. The fractional instability at 1-s sample time is found to be 4.3$\times$10$^{-13}$, closed to that for the referenced signal from the signal generator. With a longer averaging time to 1000 s, the frequency instability would approach to 8.3$\times$10$^{-15}$. The achieved instability is improved tenfold comparing to previous results based on the optical injection \cite{Yu2021OE}. Indeed, the direct use of the electrical pulse as the reference can eliminate the timing jitter during the light chopping to prepare the master optical pulse.

\begin{figure}[b!]
\centering
\includegraphics[width=0.64\columnwidth]{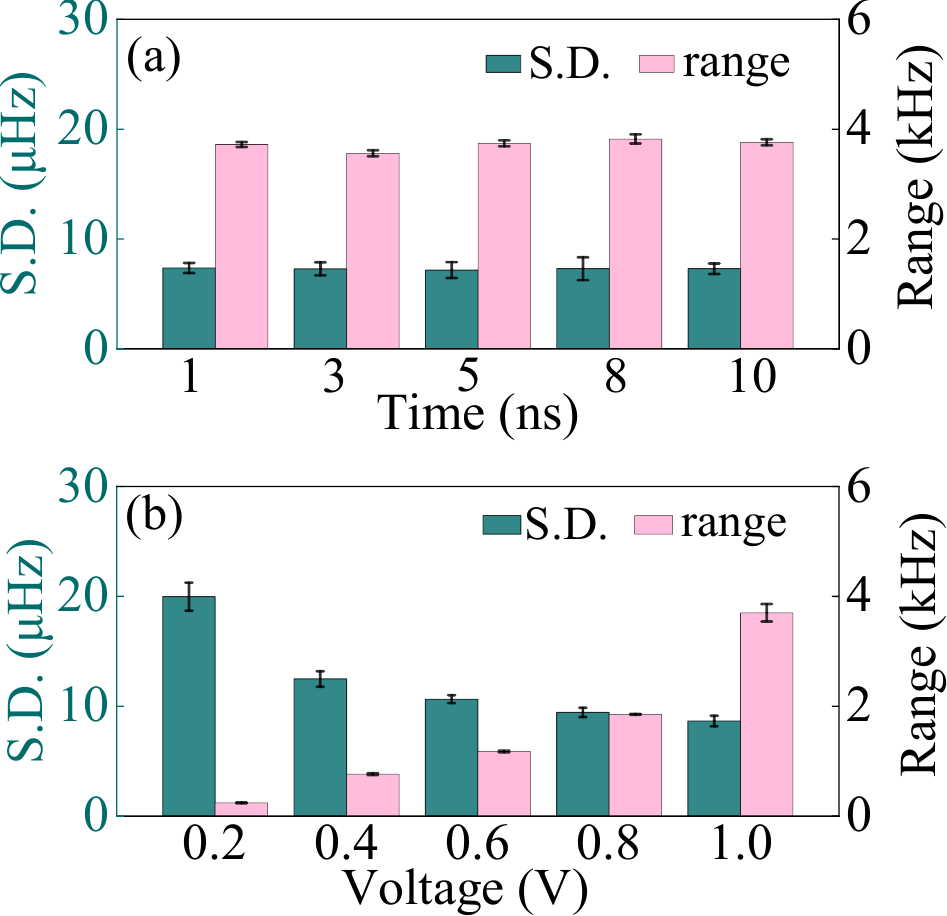}
\caption{Standard deviation (S.D.) and capture range of the repetition rate for the locked fiber laser as a function of the pulse duration (a) and voltage amplitude (b) for the electric pulse. The S.D. measurement is recorded for 1 minutes with a 1-s sample gate. The error bar on each data point is calculated from 5 sequential acquisitions.}
\label{fig4}
\end{figure}

Next, we turn to investigate the passive locking performance with a dependence on the electrical pulse parameters. As shown in Fig. \ref{fig4}(a), the stabilization precision and capture range barely change as increasing the pulse duration from 1 to 10 ns when the peak voltage here is fixed at 1 V. The optical pulse is preferably locked at a certain point on the rising edge of the electrical pulse, where the corresponding voltage should be large enough to launch the synchronous perturbation, but not too much to disrupt the mode-locking. Figure \ref{fig4}(b) presents the performance for various voltage amplitudes from 0.2 to 1 V at a fixed pulse width of 10 ns. It can be seen that a larger voltage would lead to a smaller repetition-rate fluctuation and a large dynamic range. Indeed, the resulting steeper rising edge provides a stronger pulling force in the passive synchronization.

\begin{figure}[t!]
\centering
\includegraphics[width=0.75\columnwidth]{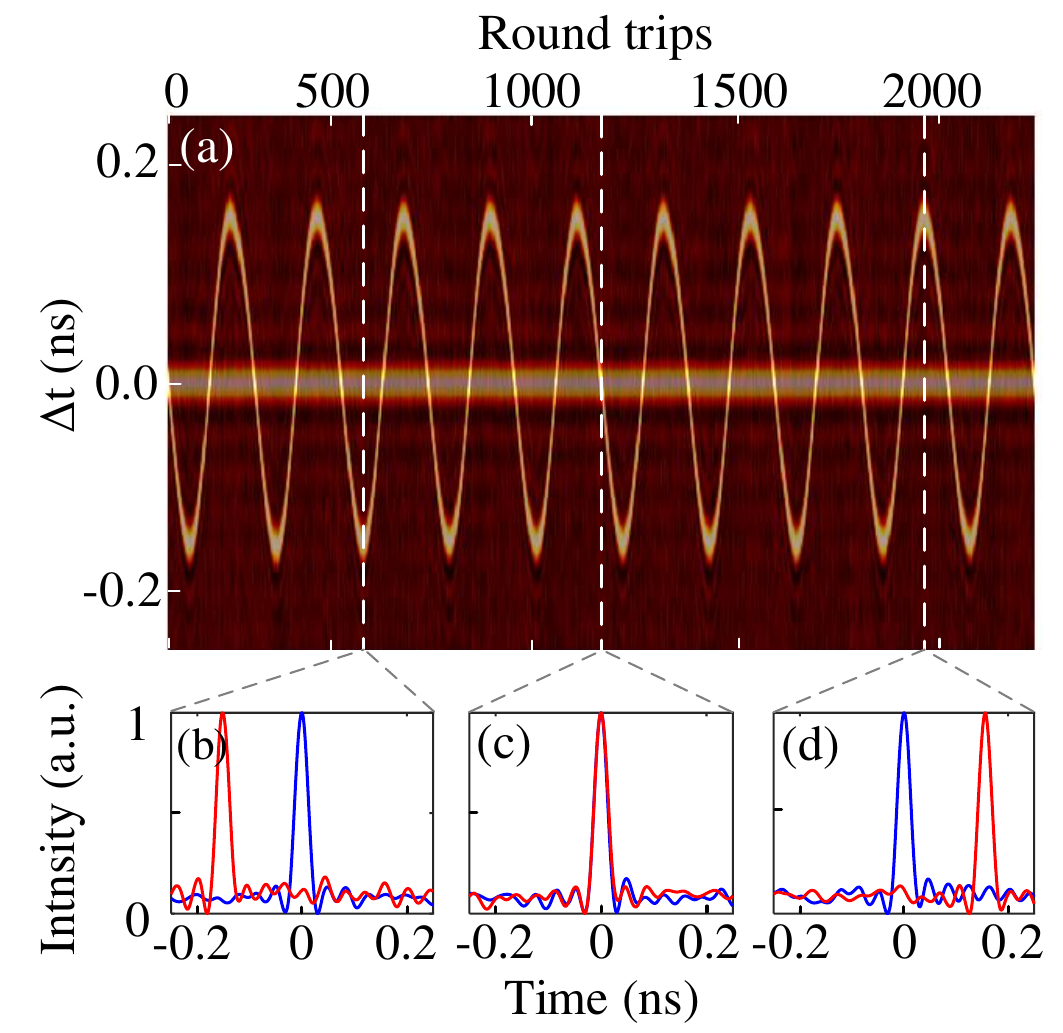}
\caption{(a) Temporal evolution of the pulses from two highly imbalanced arms in the dynamic optical sampling configuration shown in Fig. \ref{fig1}(c). One pulse is aligned at the center to highlight the relative pulse delay. Three sections in dashed lines are illustrated in (b-d). Note that each round-trip time corresponds to about 45.9 ns.}
\label{fig5}
\end{figure}

Finally, we demonstrate the repetition-rate tuning capability by modulating the driving frequency of the SG. The repetition-rate chirping is then used to implement an optical sampling between two self-delayed pulse trains as shown in Fig. \ref{fig1}(c). The two imbalanced arms differ by about 780 m. At the modulation depth of 1 kHz and rate of 100 kHz, the two pulse trains are measured by a high-speed digital oscilloscope (Agilent, DSA-X 92504A) with a bandwidth of 25 GHz and a sampling rate of 80 G/s. Consequently, a relative delay up to 305 ps for the pair-wise pulses can be scanned in 10 $\mu$s as given in Fig. \ref{fig5}. Comparing to conventional optical sampling by cavity tuning (OSCAT) \cite{Hochrein2010OE}, the non-mechanical scanning in our scheme favors a high modulation frequency and a large tuning range.  

To conclude, we have presented a passive scheme to stabilize the repetition rate for a mode-locked fiber laser via electro-optic modulation. The implemented stabilization system exhibits a large dynamic range up to 2.3 mm, which is significantly longer than previous active feedback scheme based on optical-path variation by the EOM \cite{Hudson2005OL}. The rapid passive feedback at each round trip leads to a high servo bandwidth, which facilitates a repetition-rate fractional instability as low as 4.3$\times$10$^{-13}$ in 1 s. In the proof-of-the-principle demonstration, an optical sampling by repetition-rate modulation is realized to support a 305 ps scanning range at a rate of 100 kHz. Along with the features of simple operation, compact layout, and long-term stability, the implemented laser source would favor immediate OSCAT-related applications, including distance measurement \cite{Wu2016OL}, depth-resolved imaging \cite{Yang2015PTL}, and dual-comb spectroscopy \cite{Carlson2018OL}.

\begin{backmatter}
\bmsection{Funding} National Key Research and Development Program (2021YFB2801100); National Natural Science Foundation of China (62175064, 11621404, 11727812); Fundamental Research Funds for the Central Universities; National Key Laboratory of Science and Technology on Space Microwave (HTKJ2020KL504013).

\bmsection{Disclosures} The authors declare no competing interests.

\bmsection{Data availability} The data that support the findings of this study are available from the corresponding author upon reasonable request.

\end{backmatter}

\end{document}